\documentclass[prd,twocolumn,showpacs,eqsecnum,superscriptaddress]{revtex4}

\usepackage{graphicx}
\usepackage{amsmath}
\usepackage{amssymb}
\usepackage{sistyle}
\usepackage{color}

\newcommand{\be}{\begin{equation}}
\newcommand{\ee}{\end{equation}}
\newcommand{\ba}{\begin{eqnarray}}
\newcommand{\ea}{\end{eqnarray}}

\newcount\bozza \bozza=0
\ifnum\bozza=1
\newdimen\shift \shift=-2truecm
\def\lb#1{%
{\label{#1}\rlap{\kern\shift{$\scriptstyle#1$}}}}
\else\def\lb#1{\label{#1}} \fi

\begin{document}

\title{Landau levels and magnetic oscillations in gapped Dirac materials with intrinsic Rashba interaction}

\author{V.Yu.~Tsaran}
%\thanks{On leave of absence from }
%\email{gorbar@bitp.kiev.ua}
\affiliation{Department of Physics, Taras Shevchenko National Kiev University, 6 Academician Glushkov ave.,
Kiev 03680, Ukraine}

\author{S.G.~Sharapov}
%\email{sharapov@bitp.kiev.ua}
\affiliation{Bogolyubov Institute for Theoretical Physics, National Academy of Science of Ukraine, 14-b
        Metrologicheskaya Street, Kiev 03680, Ukraine}
\affiliation{Department of Physics, Taras Shevchenko National Kiev University, 6 Academician Glushkov ave.,
Kiev 03680, Ukraine}

\date{\today }

\begin{abstract}
A new family of the low-buckled Dirac materials which includes silicene, germanene, etc. is expected
to possess a more complicated sequence of Landau levels than in pristine graphene.
Their energies depend, among other factors, on the strength of the intrinsic spin-orbit (SO) and
Rashba SO couplings and  can be tuned by an applied  electric field $E_z$.
We studied  the influence of the intrinsic Rashba SO term on the energies of Landau levels
using both analytical and numerical methods.
The quantum magnetic oscillations of the density of states are also investigated.
A specific feature of the oscillations is the presence of the beats
with the frequency proportional to the field $E_z$.
The frequency of the beats becomes also dependent on
the carrier concentration when Rashba interaction is present allowing experimental determination of its strength.
\end{abstract}

%Landau levels, 71.70.Di

% graphene, 72.80.Vp

%Graphene

%electronic transport, 72.80.Vp

%films, 68.65.Pq

%material science aspect of, 81.05.ue

% spin polarized transport in, 72.25.Dc

%\keywords{QUANTUM HALL EFFECT; GRAPHENE}

\maketitle

\section{Introduction}
\label{sec:intro}

Synthesis of silicene \cite{Lalmi2010APL,Padova2010APL,Padova2011APL,Vogt20102PRL,Lin2012APExp,Fleurence2012PRL,Chen2012PRL,Majzik2013JPCM},
a monolayer of silicon atoms forming a two-dimensional low-buckled honeycomb lattice, boosted theoretical
studies of a wide class of new buckled Dirac materials.
The honeycomb lattice of silicene can be described as in graphene in terms of two triangular sublattices.
However, a larger ionic size of silicon atoms results in the
buckling of the two-dimensional (2D) lattice. Accordingly, the sites on the two sublattices
are situated in different vertical planes with
the separation of $2 d \approx 0.46 \mbox{\AA} $.
Consequently, silicene is expected \cite{Cahangirov2009PRL,Drummond2012PRB,Liu2011PRL,Liu2011PRB}
to have a strong intrinsic spin-orbit (SO) interaction  that results in  a sizable
SO gap, $\Delta _{\text{SO}}$, in the quasiparticle spectrum opened at the Dirac points.
Moreover, by applying an electric field $E_z$ perpendicular to the plane
it possible to create  the on-site potential difference between the two sublattices
and to open also the second gap, $\Delta= E_z d$, in the quasiparticle spectrum.
Similar structure and properties are also expected in 2D sheets of Ge, Sn, P atoms
(the corresponding materials are coined as germanene, stanene and phosphorene), and Pb \cite{Tsai2013NatComm,Xu2013PRL}.

Accordingly, the charge carriers in these buckled materials have to be regarded as the gapped
Dirac fermions, in a contrast to the gapless fermions in monolayer graphene. The gap is equal to
\begin{equation}
\label{gap-valley-spin}
\Delta_{\xi \sigma} = \Delta - \xi s_\sigma \Delta _{\text{SO}},
\end{equation}
where
$\xi = \pm$ and $\sigma= \uparrow, \downarrow $ with $s_{\uparrow, \downarrow} = \pm$
are, respectively, valley and spin indices.

First principles calculations \cite{Liu2011PRL,Liu2011PRB,Tsai2013NatComm} show that the SO gap $\Delta _{\text{SO}}$
is a material dependent constant, viz.  $\Delta _{\text{SO}} \approx \SI{4.2}{meV}$ in silicene
and $\Delta _{\text{SO}} \approx \SI{11.8}{meV}$ in germanene.
On the contrary, the gap $\Delta$ is tunable in the wide range of energies $\sim \pm \SI{50}{meV}$
by varying the electric field $E_z$.
In this respect silicene and other low-buckled monolayer Dirac materials more resemble bilayer graphene.
This creates new possibilities for manipulating dispersion of electrons.
In particular, there is a prediction \cite{Ezawa2012NJP,Drummond2012PRB} that
when the gap $\Delta_{\xi \sigma}$ vanishes at $|E_z| =  E_c$ with the critical electric field
$E_c = \Delta _{\text{SO}}/d$ silicene undergoes a transition from a topological insulator (TI)
for $|E_z| < E_c$ to a band insulator (BI) for $|E_z| > E_c$.

Although silicene has already been synthesized, its exploration is still in the initial stage.
The STM and ARPES data are confirming \cite{Lalmi2010APL,Padova2010APL,Padova2011APL,Vogt20102PRL,Lin2012APExp,Fleurence2012PRL,Chen2012PRL,Majzik2013JPCM}
the main theoretical conceptions about buckled honeycomb arrangement of Si atoms and a likely
presence of the Dirac fermions near the $\mathbf{K}$ points of the Brillouin zone.
Since silicene is only available on Ag and ZrB$_2$ \cite{Fleurence2012PRL} substrates which are both conductive,
there are no yet transport and optical measurements which would ultimately confirm the Dirac nature
of the charge carriers. In general, the experimental investigations of silicene and other related  Dirac materials
are somewhat behind the theoretical ones. For example, there exist predictions
for the abovementioned transition from TI to BI \cite{Ezawa2012NJP,Drummond2012PRB}, the sequence of Landau levels \cite{Ezawa2012JPSJ,Tabert2013PRL,Tabert2013aPRB} and
density of states in an external magnetic field \cite{Tabert2013PRL,Tabert2013aPRB},
the quantum Hall \cite{Ezawa2012JPSJ} and spin Hall \cite{Dyrdal2012PSS} effects, and
optical \cite{Stille2012PRB,Matthes2013PRB,Tabert2013PRB}
and magneto-optical \cite{Tabert2013PRL,Tabert2013aPRB} conductivities.

A simple, but still capturing basic electronic
properties of silicene and other buckled Dirac materials model with the gapped Dirac fermions
was used in  most of the mentioned above studies.
The corresponding quasiparticle excitations with the gap (\ref{gap-valley-spin})
represent four (two identical pairs)  noninteracting  species of
the massive Dirac particles with  the mass $\Delta_{\xi \sigma}/v_F^2$, where $v_F$ is the Fermi velocity.
Thus the expected electronic properties of silicene \cite{Stille2012PRB,Tabert2013PRB,Tabert2013PRL,Tabert2013aPRB}
in this approximation resemble that of the two independent pieces of the gapped monolayer graphene
\cite{Gusynin2006-2007} with the gaps $\Delta \pm \Delta _{\text{SO}} $.

However, this simple picture breaks down when other interactions present in the buckled Dirac materials
are taken into account, because these interactions make the two species of the Dirac fermions with the different masses
interacting.
The most important among them is the spin-nonconserving intrinsic Rashba SO interaction, the strength $\Delta_R$
of which is defined by the coupling between second-nearest-neighboring sites \cite{Drummond2012PRB,Ezawa2012NJP}.
In an external magnetic field for $\Delta_R \neq 0$  theoretical description of silicene and other monolayer Dirac
materials by its complexity resembles treatment of a biased bilayer graphene  \cite{Mucha2009JPCM,Mucha2009SSC,Pereira2007PRB,Zhang2011PRB}
when the trigonal warping term is neglected. Although this problem is exactly solvable, there is no
explicit generic expression for the Landau level energies.

%the two-fold valley degeneracy of Landau levels  is removed.\cite{Ezawa2012JPSJ}.
%except for the points where two levels cross.\cite{Ezawa2012JPSJ}.
The purpose of the present paper is to study the influence of the parameters describing the low-buckled Dirac materials
on the energies of Landau levels and quantum  magnetic oscillations of the density of states.
The paper is organized as follows. We begin by presenting
in Sec.~\ref{sec:model} the tight binding model describing
low-buckled Dirac materials. The theory with two identical pairs of
the massive Dirac fermions is obtained in the continuum limit.
We discuss the specific of Rashba SO interaction in silicene and related materials.
In Sec.~\ref{sec:LL} we consider the structure of Landau levels.
To make the presentation thorough and consistent in Sec.~\ref{sec:LL-no-Rashba}
we begin with the overview of the results
obtained in the absence of the Rashba interaction \cite{Ezawa2012JPSJ,Tabert2013PRL,Tabert2013aPRB}.
Then in Sec.~\ref{sec:LL-Rashba} we investigate the influence of the intrinsic Rashba coupling
on the energies of Landau levels and compare our results with the existing from the paper by
Ezawa \cite{Ezawa2012JPSJ}.
In Sec.~\ref{sec:Rashba-analytic} the limiting cases that were not analyzed before are presented.
In particular, we obtain the analytic expression for energies of the Landau levels in the  quasiclassical regime.
The oscillations of the density of states (DOS) are  considered in Sec.~\ref{sec:DOS-oscillations}.
In Sec.~\ref{sec:concl}, the main results of the paper are summarized.

\section{Models and notation}
\label{sec:model}

The silicene and related Dirac materials with buckled lattice structure are described by
the four-band second-nearest-neighbor tight binding model on the honeycomb lattice \cite{Liu2011PRL,Liu2011PRB}
\begin{equation}  \label{lattice-Hamiltonian}
\begin{split}
\! \! \! \! \! \! H = & - t \sum_{ \langle i,j \rangle \sigma} \hat c^{\dagger}_{i \sigma} \hat c_{j \sigma} +
i \frac{\Delta _{\text{SO}}}{3 \sqrt{3}} \sum_{\substack{\langle \!\langle i,j \rangle \!\rangle\\ \sigma \sigma^\prime }}
 \hat c^{\dagger}_{i \sigma} (\pmb{\nu}_{i j} \cdot \pmb{\sigma})_{\sigma \sigma^\prime} \hat c_{j \sigma^\prime} \\
& - i \frac{2}{3} \Delta_R \sum_{\substack{\langle \! \langle i,j \rangle \! \rangle \\ \sigma \sigma^\prime}} \mu_{ij}
\hat c^{\dagger}_{i \sigma} (\pmb{\sigma} \times \hat{\mathbf{f}}_{ij})^{z}_{\sigma \sigma^\prime} \hat c_{j \sigma^\prime} \\
&+ \sum_{i \sigma} ( \eta_i \Delta -\mu) \hat c^{\dagger}_{i \sigma} \hat c_{i \sigma},
\end{split}%
\end{equation}
where $\hat c^{\dagger}_{i \sigma}$ creates an electron on a site $i$ with spin $\sigma$.
The sum is taken over all pairs of nearest-neighbour (NN)
and next-nearest-neighbor (NNN) lattice sites that are denoted, respectively, by the
symbols $ \langle i,j \rangle$ and $\langle \!\langle i,j \rangle \! \rangle$ which also implicitly include
the Hermitian conjugate terms.

The honeycomb lattice with a lattice constant $a$
consists of two $A$ and $B$ sublattices, and is spanned by the basis vectors
$\mathbf{a}_1= a(\frac{1}{2}, \frac{\sqrt{3}}{2})$ and  $\mathbf{a}_2=a(\frac{%
1}{2},-\frac{\sqrt{3}}{2})$. The lattice constant
$a = |\mathbf{a}_1| = |\mathbf{a}_2| = \sqrt{3} a_{NN}$ and $a_{NN}$ is the distance between two NN
atoms.

The first term in (\ref{lattice-Hamiltonian}) is the usual tight-binding NN hopping
between sites on different sublattices with the transfer energy $t$ which
results in the well-known band structure of graphene.
The second term is the intrinsic SO interaction with the coupling $\Delta_{\text{SO}}$
described by complex-valued NNN hopping with a sign $\pm 1$ which depends on the sublattice,
the direction of the hop (i.e.б clockwise or anticlockwise), and spin orientation. This sign is encoded
in $\pmb{\nu}_{i j} \cdot \pmb{\sigma}$, where the vector $\pmb{\sigma} = (\sigma_1, \sigma_2, \sigma_3)$ is made of
the Pauli spin matrices
and
\begin{equation}
\pmb{\nu}_{i j} = \frac{\mathbf{d}_{ik} \times \mathbf{d}_{kj}}{|\mathbf{d}_{ik} \times \mathbf{d}_{kj} |},
\end{equation}
with $\mathbf{d}_{ik}$ being the vector connecting NN sites $i$ and $k$, and $k$  the intermediate lattice site
involved in the hopping process from site $i$ to NNN site $j$.
The third term represents the intrinsic Rashba SO interaction with the coupling constant
$\Delta_R$, where $\mu_{ij} = 1 (-1)$ when linking the $A - A$ ($B - B$) sites,
$\mathbf{f}_{ij}$ is connecting the NNN sites, and $\hat{\mathbf{f}}_{ij}= \mathbf{f}_{ij}/|\mathbf{f}_{ij}|$.
The fourth term, which breaks the inversion symmetry, involves the staggered sublattice potential $\Delta$,
where $\eta_i = \pm 1$ for the $A$ $(B)$ site. It arises  when the external electric field $E_z$ is applied.
The chemical potential $\mu$ is also included in the Hamiltonian.

The Hamiltonian which includes the first three terms of (\ref{lattice-Hamiltonian}) is also
called the Kane-Mele Hubbard model \cite{Hohenadler2013JPCM} and was originally proposed
as a model for graphene \cite{Kane2005PRL}. In the present case, however, both the intrinsic SO and Rashba
SO terms originate from buckling of the lattice structure and thus
distinguish silicene, germanene, stanene, and other similar materials
from graphene where these terms are negligibly small. Both the intrinsic SO and Rashba SO
terms respect the inversion symmetry, but the Rashba term breaks $z \to - z$ symmetry. The Rashba term
is purely off-diagonal in spin, so its presence makes the spin nonconserving.
The Hamiltonian (\ref{lattice-Hamiltonian}) respects also the time-reversal symmetry.

It is important to stress that the intrinsic Rashba term included in Eq.~(\ref{lattice-Hamiltonian})
involves hopping between NNN sites, while in the Kane-Mele model the hopping is between NN sites.
As in graphene \cite{Kane2005PRL}, in silicene and related materials the NN Rashba term is extrinsic.
It may be induced by the external electric field $E_z$ or by interaction with a substrate \cite{Varykhalov2008PRL}.
The influence of the extrinsic NN Rashba term on the spectrum of Landau levels was studied in  \cite{Rashba2009PRB,Martino2011PRB}.
In the present paper we do not consider interaction with a substrate that potentially can affect not only this specific term,
but also other terms of the Hamiltonian (\ref{lattice-Hamiltonian}). We focus only on the role of the electric field $E_z$
allowing the gap $\Delta$ to be a free parameter of the model.
In this case, as discussed in \cite{Ezawa2012PRL}, the extrinsic  Rashba term can be safely neglected, because
it is two or three orders of magnitude less than the intrinsic  Rashba term.

Near two independent $\mathbf{K}_{\pm} = \pm 2 \pi/a(2/3,0)$ points the bare band dispersion
provided by the first term of (\ref{lattice-Hamiltonian}) is linear, $\varepsilon(\mathbf{k}) = \pm \hbar v_F k$,
with the Fermi velocity
$v_F=\sqrt{3}t a /(2 \hbar)$. Its theoretical estimates (see, e.g., Refs.~\cite{Drummond2012PRB,Liu2011PRL,Liu2011PRB,Tsai2013NatComm,Xu2013PRL})
gave the value $v_F \sim 5 \times 10^5 \mbox{m/s}$ for all family of the new materials, while
measurements done in silicene \cite{Vogt20102PRL,Chen2012PRL} suggest that $v_F \sim 10^6 \mbox{m/s}$
which is close to the observed in graphene.

There are no reliable data for the value $\Delta _{\text{SO}}$, but as mentioned in the Introduction,
its theoretical estimates \cite{Liu2011PRL,Liu2011PRB,Tsai2013NatComm} give the value $\Delta _{\text{SO}}$
order of \SI{10}{meV} in silicene and germanene, and even
$\sim 100 - 200 \, \mbox{meV} $ in Sn and Pb. The same papers provide the estimates for
$\Delta_R  \sim 1 - 20 \, \mbox{meV} $.

The physics of conducting electrons in silicene and other buckled materials can be successfully described
by the low-energy Dirac theory. In the simplest case of graphene it is enough to include only the first and the last $\sim \mu$
terms of the lattice Hamiltonian  (\ref{lattice-Hamiltonian}) which result in the massless QED$_{2+1}$
effective theory with four (two valleys and two spins) identical flavours of fermions. A more involved case of
silicene requires that the other terms of the Hamiltonian
(\ref{lattice-Hamiltonian}) have to be taken into account.
The resulting low-energy Hamiltonian in the momentum representation reads
\begin{equation}
\label{Hamiltonian-momentum}
H = \sum_{\xi = \pm} \int \frac{d^2 k}{(2 \pi)^2} \Psi^\dagger_\xi(\mathbf{k}) \mathcal{H}_\xi(\mathbf{k}) \Psi_\xi (\mathbf{k}),
\end{equation}
where $\xi = \pm$ at $\mathbf{K}_{\pm}$ points (valleys) and
\begin{equation}
\label{basis}
%\begin{split}
\Psi_\xi(\mathbf{k})=
\left(
  \begin{array}{c}
    \psi_{A\uparrow} (\mathbf{K}_\xi + \mathbf{k} ) \\
    \psi_{B\uparrow} (\mathbf{K}_\xi + \mathbf{k} ) \\
    \psi_{A\downarrow} (\mathbf{K}_\xi + \mathbf{k} ) \\
    \psi_{B\downarrow} (\mathbf{K}_\xi + \mathbf{k} ) \\
  \end{array}
\right)
%\end{split}
\end{equation}
is the spinor made from the Fermi operators $\psi_{A\sigma} (\mathbf{K}_\xi + \mathbf{k})$,
$\psi_{B\sigma} (\mathbf{K}_\xi + \mathbf{k})$
of electrons  on $A$ and $B$ sublattices with spin $\sigma$ and the wave-vector $\mathbf{k}$
measured from the $\mathbf{K}_\pm$ points. The Hamiltonian density for $\mathbf{K}_{\pm}$ points
is $\mathcal{H}_\xi(\mathbf{k}) =\mathcal{H}_\xi^0(\mathbf{k}) + \mathcal{H}_\xi^R(\mathbf{k})$
with
\begin{equation}
\label{Hamiltonian-density-0}
\begin{split}
\mathcal{H}_\xi^0(\mathbf{k})  = & \sigma_0 \otimes [\hbar v_F (\xi k_x \tau_1 + k_y\tau_2 ) + \Delta \tau_3 - \mu \tau_0]  \\
& - \xi \Delta_{\mathrm{SO}} \sigma_3 \otimes \tau_3
\end{split}
\end{equation}
and
\begin{equation}
\label{Hamiltonian-density-R}
\mathcal{H}_\xi^R(\mathbf{k})=- a\Delta_R(k_y \sigma_1-k_x\sigma_2) \otimes \tau_3.
\end{equation}
Here the Pauli matrices $\pmb{\tau}$ act in the sublattice space and as above the matrices
$\pmb{\sigma}$ act in the spin space, $\tau_0$ and $\sigma_0$ are the unit matrices.
The Hamiltonian density (\ref{Hamiltonian-density-0}) describes
noninteracting massive Dirac quasiparticles with the gaps (masses) $\Delta_{\xi \sigma}$ given by Eq.~(\ref{gap-valley-spin}).
The presence of the mass term reduces the fourfold degeneracy between fermion flavors to the twofold degeneracy.
Moreover, the Rashba term (\ref{Hamiltonian-density-R}) introduces interaction between
fermions with the opposite spin within each valley.
Notice that the NNN character of the Rashba term results in the presence of the wave vector $\mathbf{k}$
in Eq.~(\ref{Hamiltonian-density-R}) and in our conventions it turns out to be the  same for both $\mathbf{K}_{\pm}$
points.

One can verify that the Hamiltonian $\mathcal{H}(\mathbf{k}) = \mathcal{H}_{\xi=+1}(\mathbf{k}) \oplus \mathcal{H}_{\xi=-1}(\mathbf{k}) $
respects time-reversal symmetry that in the basis we use is described by \cite{Gusynin2007IJMPB}
\begin{equation}
(\Pi \otimes \sigma_2 \otimes \tau_0 ) \mathcal{H}^\ast(\mathbf{k}) (\Pi \otimes \sigma_2 \otimes \tau_0 ) =
\mathcal{H}(-\mathbf{k}).
\end{equation}
Here $\Pi$ swaps $\xi=1$ and $\xi =-1$ valleys.
The SO and Rashba SO terms in the continuum Hamiltonian (\ref{Hamiltonian-momentum}),
(\ref{Hamiltonian-density-0}) and (\ref{Hamiltonian-density-R})
also respect the inversion symmetry that exchanges both the sublattices and $\mathbf K_{\pm}$
points.

It is convenient to redefine the spinor $\Psi_-(\mathbf{k})$ at $\mathbf{K}_-$ point
by swapping sublattices in its $\sigma = \downarrow$ part
\begin{equation}
\label{basis-swapped}
\Psi_-(\mathbf{k})=
\left(
  \begin{array}{c}
    \psi_{A\uparrow} (\mathbf{K}_- + \mathbf{k} ) \\
    \psi_{B\uparrow} (\mathbf{K}_- + \mathbf{k} ) \\
    \psi_{B\downarrow} (\mathbf{K}_- + \mathbf{k} ) \\
    \psi_{A\downarrow} (\mathbf{K}_- + \mathbf{k} ) \\
  \end{array}
\right).
%\end{split}
\end{equation}
Then the Hamiltonian density for both  $\mathbf{K}_\pm$ points can be
combined in one matrix
\begin{equation}
\label{Hamiltonian-generic-B=0}
\mathcal{H}_{\xi}=\xi
\begin{pmatrix}
\Delta_{\xi \uparrow}  &   \hbar v_F k_-  &  - i a\Delta_R k_-   &  0\\
\hbar v_F k_+ & -\Delta_{\xi\uparrow} &  0    &  i a \Delta_R k_-\\
 i a \Delta_R k_+   &  0&  \Delta_{\xi \downarrow}   &   \hbar v_F k_-\\
0  &   - i a \Delta_R k_+   &     \hbar v_F k_+  & -\Delta_{\xi \downarrow}
\end{pmatrix},
\end{equation}
where $k_\pm=k_x\pm i k_y$.
The energy spectrum of silicene in zero magnetic field \cite{Drummond2012PRB,Ezawa2012NJP,Ezawa2012JPSJ}
directly follows from Eq.~(\ref{Hamiltonian-generic-B=0})
\begin{equation}
\label{spectrum-B=0}
\epsilon_{\xi \sigma}^{\pm}= \pm \sqrt{\hbar^2 v_F^2 k^2+(\Delta- \xi s_\sigma \sqrt{\Delta_{\text{SO}}^2+a^2\Delta_R^2k^2})^2}.
\end{equation}
We observe that $\Delta_R$ appears only in the combination $a \Delta_R k$ in the spectrum (\ref{spectrum-B=0})
which vanishes at the $K_{\pm}$ points. Nevertheless, because the Rashba term is  spin nonconserving
it is important to study how its value can be extracted from the observable quantities.

\section{Landau levels}
\label{sec:LL}

In an external magnetic field $\mathbf{B} = \mathbf{\nabla} \times \mathbf{A} = (0,0,B)$ applied perpendicular
to the plane along the positive $z$ axis the momentum operator $\hbar k_i$ has to be replaced by the covariant
momentum $\hbar k_i \to \Pi_i = \hbar k_i + \frac{e}{c} A_i$. Here $-e<0$ is the electron charge
and the vector potential in the Landau gauge $\mathbf{A} = (0, Bx,0)$. Introducing a pair of Landau level
ladder operators satisfying $[\hat a,\hat a^\dag]=1$,
\begin{equation}
\label{a-def}
\hat a=\frac{l_B}{\sqrt2\hbar}(\Pi_x-i \Pi_y), \qquad
\hat a^\dag=\frac{l_B}{\sqrt2\hbar}(\Pi_x+ i\Pi_y)
\end{equation}
where $l_B = \sqrt{\hbar c/ eB}$ is the magnetic length, we rewrite
the Hamiltonian (\ref{Hamiltonian-generic-B=0}) in the form
\begin{equation}
\label{Hamiltonian-generic-B}
\mathcal{H}_\xi=\xi
\begin{pmatrix}
\Delta_{\xi\uparrow}   & \hbar\omega \hat a  & - i R\hat a  &  0\\
  \hbar\omega \hat a^\dag  & -\Delta_{\xi\uparrow} &  0    &  i R \hat a \\
 i R \hat a^\dag   &  0&  \Delta_{\xi\downarrow}    &    \hbar\omega \hat a\\
0  &   - i R \hat a^\dag  &   \hbar\omega \hat a^\dag &   -\Delta_{\xi\downarrow}
\end{pmatrix}.
\end{equation}
Here
\begin{equation}
\label{Landau-scale}
\hbar \omega = \sqrt{2}\frac{ \hbar v_F}{l_B} \approx 36.28\, v_F[\times 10^6 \mbox{m/s}] \sqrt{B [\mbox{T}]} \,\mbox{meV}
\end{equation}
is the Landau scale and
we introduced a shorthand notation for the Rashba term, $R=\sqrt{2}\frac {a}{l_B}\Delta_R$.
Note that inversion of the field direction results in the exchange of the spectra for the $\mathbf{K}_\pm$ points.

\subsection{Landau levels in the absence of Rashba term}
\label{sec:LL-no-Rashba}

When the Rashba term is absent, $R=0$, the Hamiltonian becomes block diagonal with each block
corresponding to the different direction of the spin. Within each block the Hamiltonian is identical
to that of the gapped graphene \cite{Semenoff1984PRL,Haldane1988PRL}, although the value of the gap
(\ref{gap-valley-spin})
is now spin dependent. The corresponding eigenstates represent a mixture of two states from two different
Landau levels and the  energies of the Landau levels are
\cite{Tabert2013PRL,Tabert2013aPRB}
\begin{equation}
\label{LL-(R=0)}
\begin{cases}
\epsilon_{0\xi \sigma} & =-\xi \Delta_{\xi \sigma}, \qquad \qquad \qquad  n=0, \\
\epsilon_{n\xi \sigma}^{\pm} & =
\pm \sqrt{\Delta^2_{\xi \sigma}+ n (\hbar \omega)^2}, \quad n= 1,  2, \dots
\end{cases}
\end{equation}
The energies of the four $n=0$ Landau levels do not depend on the value of the magnetic field.

\begin{figure}[ht]
\includegraphics[width=8cm]{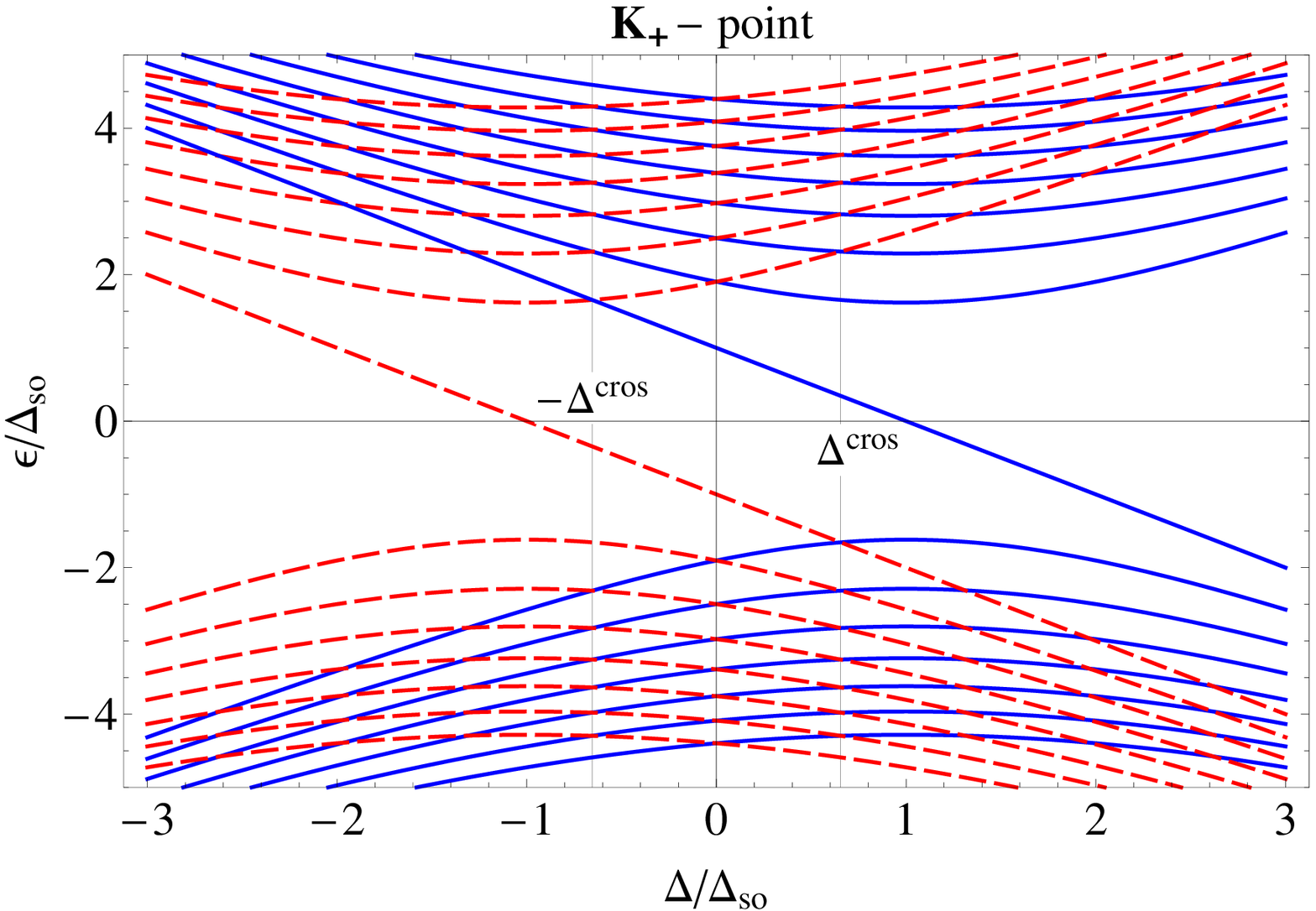}
\includegraphics[width=8cm]{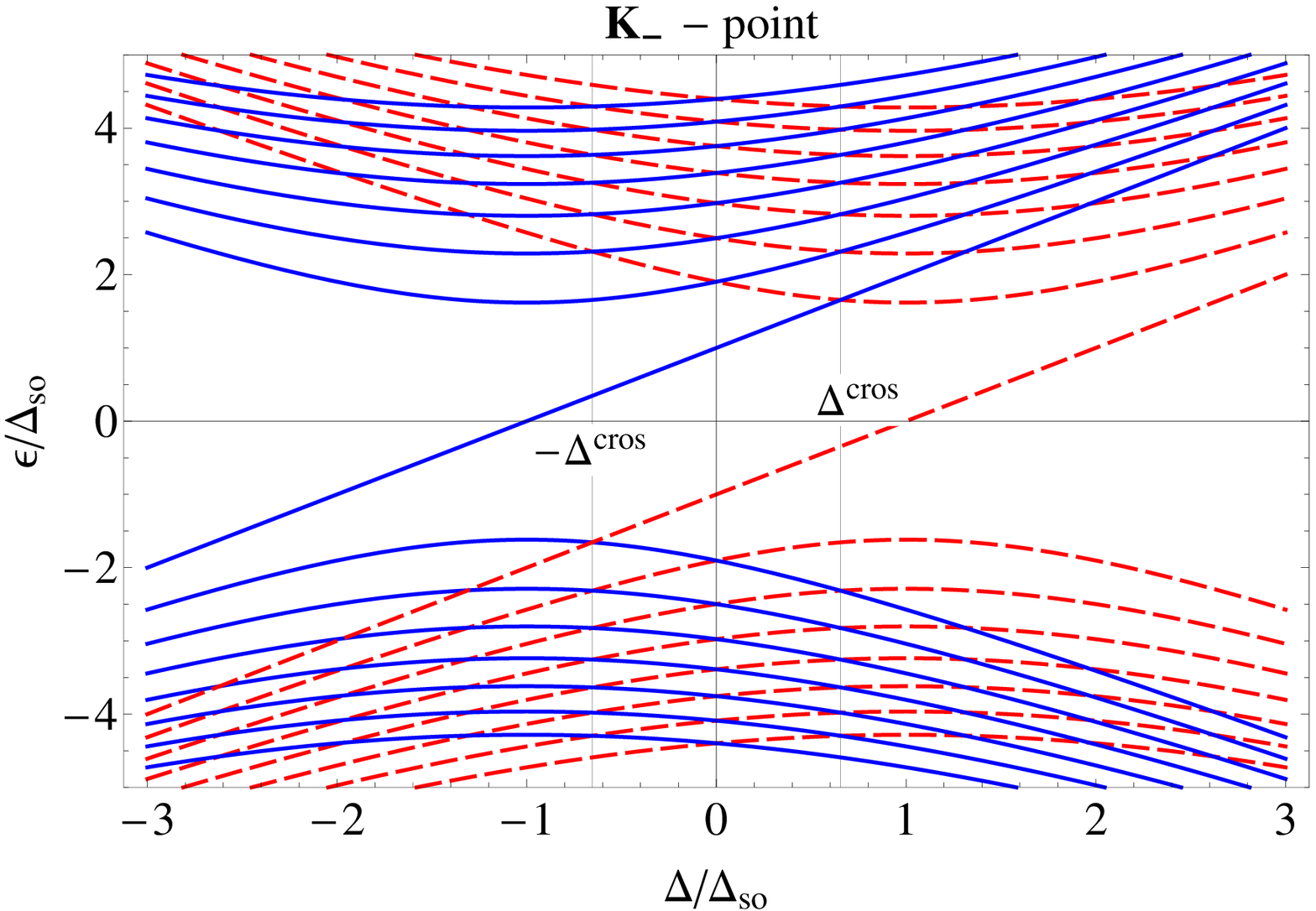}
\caption{ (Color online)
Energies of Landau levels in $\Delta _{\text{SO}}$ units at the $\mathbf{K}_{\pm}$ points (upper and lower panels, respectively)
as a function of $\Delta/\Delta _{\text{SO}}$ for  $\Delta_R=0$ and   $\hbar \omega = 1.62 \Delta _{\text{SO}}$.
The solid (blue) and dashed (red) curves correspond to $\sigma = \uparrow, \downarrow$, respectively.}
\label{fig:1}
\end{figure}

In Fig.~\ref{fig:1} we show the energies $\epsilon$ of Landau levels as a function of the sublattice asymmetry gap
$\Delta$ for $R=0$ (upper and lower panels correspond to the $\mathbf{K}_{\pm}$ points).
As was mentioned in the Introduction, the value of $\Delta$ is proportional to the electric field $E_z$ applied perpendicular
to the plane.
Here the levels with $\sigma = \uparrow$ and $\sigma= \downarrow$ are shown by the solid
(blue) and dashed (red) curves, respectively. The value of the Fermi velocity $v_F$ is assumed to be typical
for silicene.
The transition from a TI to a BI occurs at the critical value of the
gap, $|\Delta| =  \Delta_c$, where
$\Delta_{c} = E_c d = \Delta _{\text{SO}}$. Indeed, as one can see in Fig.~\ref{fig:1}
for $|\Delta| < \Delta_c$ in the TI regime the $n=0$ spin up Landau levels at the $\mathbf{K}_{\pm}$ points
have a positive energy, while the corresponding spin down levels have a negative energy.
In the BI regime for $|\Delta| > \Delta_c$ the signs of the energies of the
$n=0$ spin up levels at the $\mathbf{K}_{\pm}$ points are opposite.
The points of adjacent  level (with $n-n^\prime = \pm1$ and $\sigma^\prime = - \sigma$) crossing at
$\Delta = \pm \Delta^{\mathrm{cros}}$ are marked in Fig.~\ref{fig:1} by the vertical lines.

Concluding the overview of the results with zero Rashba term, we note that there are other Dirac materials
such as MoS$_2$ that despite having a different atomic structure are still described by
the Hamiltonian (\ref{Hamiltonian-density-0}) with $\Delta _{\text{SO}} \ll |\Delta|$ in the low-energy approximation.
Accordingly, the spectrum of Landau levels in MoS$_2$ \cite{Rose2013PRB} is
the same as considered above.

\subsection{Landau levels in the presence of Rashba term}
\label{sec:LL-Rashba}

Now we consider the eigenstates $\Psi^n_{\xi}$ of  the Hamiltonian (\ref{Hamiltonian-generic-B}) satisfying
$\mathcal{H}_\xi \Psi^n_{\xi} = \epsilon \Psi^n_{\xi}$ for the $R \neq 0$ case \cite{Ezawa2012JPSJ}.
For $n \geq 1$ these eigenstates  represent the mixture of four states from three different Landau levels
\begin{equation}
\label{eigen-n>1}
\Psi^n_{\xi} =(u_{{\xi} \uparrow}^{n-1}|n-1\rangle,  v_{{\xi} \uparrow}^{n}|n\rangle, u_{{\xi} \downarrow}^{n}|n\rangle, v_{{\xi} \downarrow}^{n+1}|n+1\rangle)^T
\end{equation}
with $|n\rangle=\frac1{\sqrt{n!}}(\hat a^\dag)^n|0\rangle$.
Then the coefficients $u_{{\xi} s}^{n}$ are the characteristic vectors of the matrix
\begin{equation}
\label{Hamiltonian-matrix-B}
\mathcal{H}_{\xi}^n= \xi
\begin{pmatrix}
\Delta_{\xi\uparrow}   & \sqrt{n} \hbar\omega  & - i \sqrt{n}R   &  0\\
\sqrt{n} \hbar\omega  &- \Delta_{\xi\uparrow}    &  0    &  i\sqrt{n+1}R\\
 i\sqrt{n}R   &  0&  \Delta_{\xi\downarrow} &   \sqrt{n+1} \hbar\omega\\
0  &   -  i \sqrt{n+1} R  &  \sqrt{n+1} \hbar\omega &   - \Delta_{\xi\downarrow}
\end{pmatrix}.
\end{equation}
The corresponding eigenenergies are found from the characteristic quartic equation
$\det ( \mathcal{H}_{\xi}^n-\epsilon \hat I)=0$ which takes the form
\begin{equation}
\label{f(E,R)=0}
\begin{split}
& \epsilon ^4- \left[2(\Delta_{\text{SO}}^2+ \Delta ^2) +(2 n+1)\hbar ^2 \tilde\omega^2 \right]\epsilon^2\\
& +\left(\Delta_{\xi\uparrow}^2+n\hbar ^2 \tilde\omega^2\right) \left(\Delta_{\xi\downarrow}^2+(n+1) \hbar ^2 \tilde\omega^2 \right) \\
 & +2R^2 \Delta  \left[\xi(\epsilon+\Delta_{\text{SO}})-(2n+1)\Delta\right]=0,
\end{split}
\end{equation}
where
\begin{equation}
\label{renormalized-scale}
\hbar \tilde\omega= \sqrt{\hbar ^2 \omega^2+R^2}
\end{equation}
is the Landau scale renormalized by the Rashba term.
Writing Eq.~(\ref{f(E,R)=0}) we explicitly isolated the last $\sim R^2 \Delta$ term; its absence would make this
equation biquadratic.

For $n=-1$ the eigenstate is
\begin{equation}
\label{eigen-n=-1}
\Psi^{-1}_{\xi}=(0,  0, 0, v_{\xi\downarrow}^{0}|0\rangle)^T,
\end{equation}
with the energy
\begin{equation}
\label{LLL-energy}
\epsilon_{0\xi\downarrow}=-\xi \Delta_{\xi\downarrow}.
\end{equation}
This solution is represented by the dashed (red) straight lines in Fig.~\ref{fig:1}.
Thus for $R \neq 0$ only two out of the four energy levels from (\ref{LL-(R=0)}) with $n=0$
that include only one state from the lowest Landau level $|0\rangle$
remain independent of the strength of the magnetic field.

For $n=0$ the eigenstate represents the mixture of three states from two different Landau levels
\begin{equation}
\label{eigen-n=0}
\Psi^{0}_{\xi}=(0,  v_{\xi\uparrow}^{0}|0\rangle, u_{\xi\downarrow}^{0}|0\rangle, v_{\xi\downarrow}^{1}|1\rangle)^T.
\end{equation}
The corresponding third order characteristic equation has the form
\begin{equation}
\label{low f(E,R)=0}
\begin{split}
& \epsilon^3+\xi \Delta_{\xi\uparrow} \epsilon^2-(\hbar^2 \tilde \omega^2+\Delta_{\xi\downarrow}^2)\epsilon \\
& -\xi \Delta_{\xi\uparrow} (\hbar^2 \tilde \omega^2+\Delta_{\xi\downarrow}^2)+2\xi \Delta R^2=0.
\end{split}
\end{equation}
In the $R=0$  limit the first root of Eq.~(\ref{low f(E,R)=0}) is
\begin{equation}
\label{cubic-root1}
\epsilon_{0\xi \uparrow}=-\xi \Delta_{\xi\uparrow}.
\end{equation}
One can see that it corresponds to the other two $n=0$ levels given by Eq.~(\ref{LL-(R=0)})
which are shown in Fig.~\ref{fig:1} as  the solid (blue) straight lines.
The other two roots for $R=0$ are
\begin{equation}
\label{cubic-root23}
 \epsilon_{1\xi \downarrow}^{\pm}=\pm\sqrt{\Delta_{\xi \downarrow}^2+\hbar^2\omega^2}
\end{equation}
which reproduces two (out of four) of the $n=1$ Landau levels from Eq.~(\ref{LL-(R=0)}).

As we already mentioned the description of monolayer low-buckled Dirac materials
resembles the formalism developed for a biased bilayer graphene \cite{Mucha2009JPCM,Mucha2009SSC,Pereira2007PRB,Zhang2011PRB}.
The former case is simpler, because the analytically solvable model Hamiltonian (\ref{Hamiltonian-generic-B})
contains all relevant parameters. In contrast, to capture the physics of bilayer graphene
it is necessary to include the trigonal warping term, presence of which makes the analysis of the Landau level spectra
totally numerical.

\begin{figure}[ht]
\includegraphics[width=8cm]{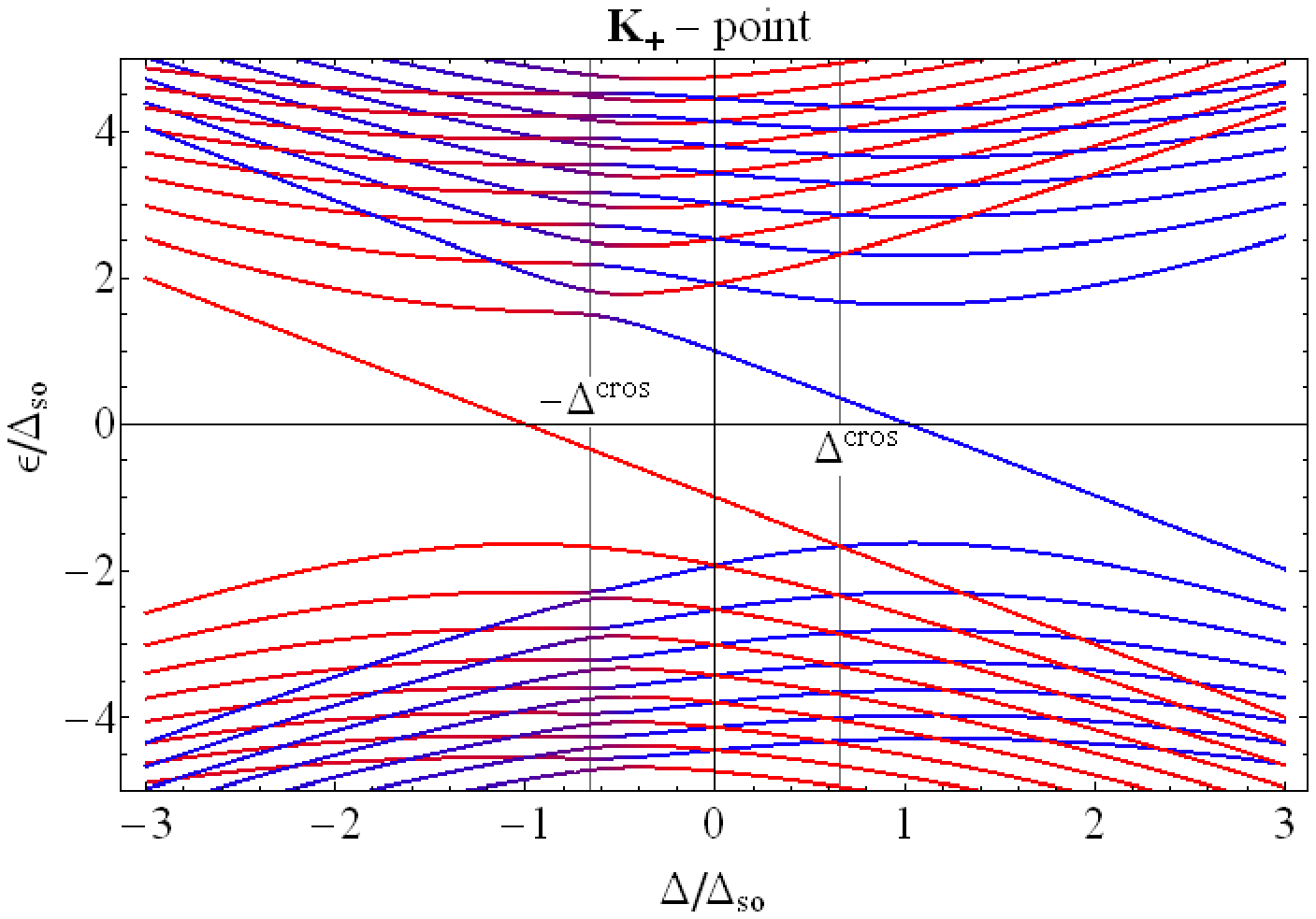}
\includegraphics[width=8cm]{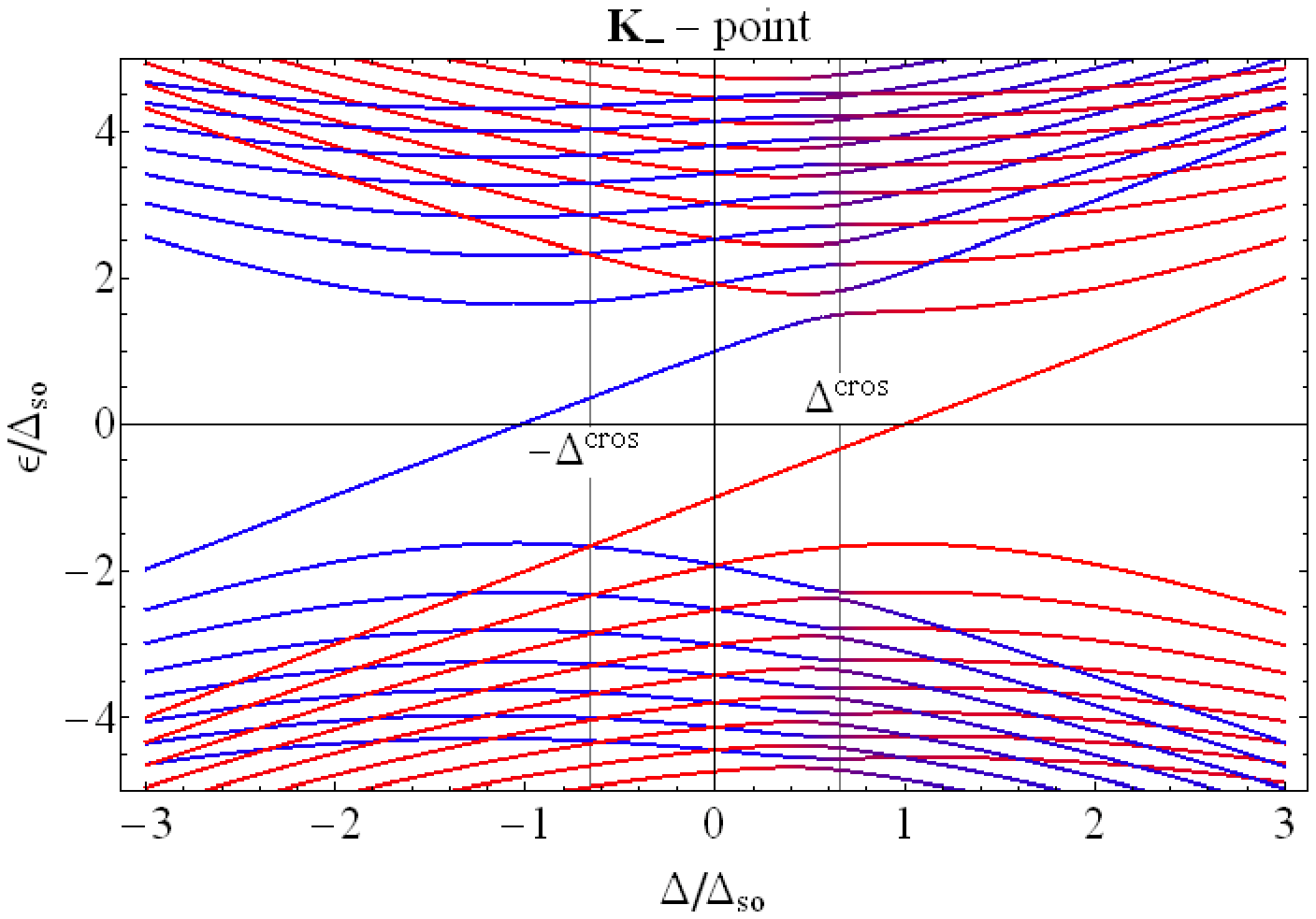}
\caption{ (Color online)
Energies of Landau levels in $\Delta _{\text{SO}}$ units at the $\mathbf{K}_{\pm}$ points (upper and lower panels, respectively)
as a function of $\Delta/\Delta _{\text{SO}}$ for $\Delta_R= 150 \Delta _{\text{SO}}$ and $\hbar \omega =1.62 \Delta _{\text{SO}}$.
The color marking corresponds to the not conserving spin quantum number: blue and red correspond to $\sigma = \uparrow, \downarrow$, respectively, and their mixture reflects the superposition of the spins.}
\label{fig:2}
\end{figure}

In spite of a  possibility to solve the quartic equation (\ref{f(E,R)=0}) and cubic equation
(\ref{low f(E,R)=0}) analytically, such general solutions are not particularly useful due to their complexity.
Thus we present in Fig.~\ref{fig:2}  the numerical solutions of Eqs.~(\ref{f(E,R)=0}) and (\ref{low f(E,R)=0}).
The same values of the parameters as in Fig.~\ref{fig:1} are taken, but for a better readability
of the figure we used the exaggerated value of the Rashba coupling constant, $\Delta_R= 150 \Delta _{\text{SO}}$.
Although for $\Delta_R \neq 0$ spin is not conserving quantum number, it is instructive to extend the color marking scheme
of Fig.~\ref{fig:1} and represent the  exact spin $|\!\uparrow\,\rangle$ and  $|\!\downarrow\,\rangle$ states
by the blue and red colors, respectively, and mark the superposition spin
state $|\alpha\rangle=a |\!\uparrow\,\rangle+b |\!\downarrow\,\rangle$
by the mixture of the colors in RGB space $\{|b|^2,0,|a|^2\}$. We observe that for $|\Delta| \gg \Delta _{\text{SO}}$
the spin states remain practically pure and their mixing occurs only in the vicinity of the anticrossing points.
At these points $a=b=1/2$.

Figure~\ref{fig:2} is similar to but not identical with the corresponding
figure from the paper of Ezawa \cite{Ezawa2012JPSJ}, because the continuum model
(\ref{Hamiltonian-generic-B=0}) has some sign difference.
In particular, we observe that in Fig.~\ref{fig:2} for the $\mathbf{K}_{+}$ point the adjacent level ($n-n^\prime = \pm1$ and
$\sigma^\prime = - \sigma$) anticrossing occurs at $\Delta = - \Delta^{\mathrm{cros}}$, while
for the  $\mathbf{K}_{-}$ point the adjacent level anticrossing takes place at $\Delta =  \Delta^{\mathrm{cros}}$.
At first glance, the influence of the Rashba term on the spectrum of the low lying Landau levels
turns out to be essential only near the anticrossing points, while outside of these regions the pattern
of the Landau levels remains almost unchanged as compared to Fig.~\ref{fig:1}.
A careful analysis shows that the situation is more complicated.

\begin{figure}[ht]
\includegraphics[width=8cm]{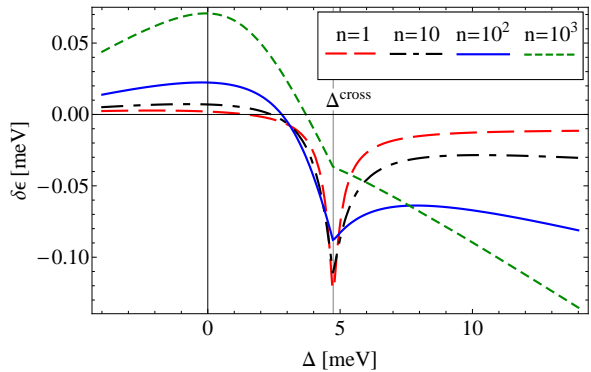}
\caption{ (Color online)
The dependence of the energy difference $\delta\epsilon=\epsilon_{n \xi}(R\ne0,\Delta)- \epsilon_{n \xi}(R=0,\Delta)$
in $\text{meV}$  as a function of  $\Delta$ in $\text{meV}$ for $n=1,10, 10^2,10^3$ at the $\mathbf{K}_{-}$ point,
$B= \SI{0.2}{T}$, $\Delta_{\text{SO}}= \SI{4.2}{meV}$, and $\Delta_R= \SI{21}{meV}$. The Fermi velocity $v_F = 5.5 \times 10^5 \mbox{m/s}$.}
\label{fig:3}
\end{figure}

To look closer at the impact of the Rashba term, in Fig.~\ref{fig:3} we plotted the dependence of the energy difference
$\delta\epsilon=\epsilon_{n \xi}(R\ne0,\Delta)- \epsilon_{n \xi}(R=0,\Delta)$
on  the value of the gap $\Delta$  at the $\mathbf{K}_{-}$ point  for four different Landau levels.
The long dashed (red) curve is for $n=1$, the dash–dotted (black) curve is for $n=10$,
solid (blue) curve is for $n=10^2$, and the short dashed (green) is for $n=10^3$.
We observe that for small $n=1 - 10$ the Rashba term is indeed important only near the anticrossing
point, $\Delta = - \xi \Delta^{\mathrm{cros}}$. Moreover, its impact on the spectrum at
the point $\Delta = - \xi \Delta^{\mathrm{cros}}$ is stronger for smaller $n$.
On the other hand, for large $n \gtrsim 100$ the Rashba term becomes more important
for $|\Delta| \gtrsim  \Delta^{\mathrm{cros}}$.

Now we proceed to the discussion of the cases where the analytical work can provide more
insight to the role of the Rashba term.

\subsection{Analytical treatment of the Rashba term }
\label{sec:Rashba-analytic}

The analytical consideration of the Rashba term may be
useful in the following cases.\\
(i) One can obtain a simple generalization of the spectrum
(\ref{LL-(R=0)}) which would allow one to consider the influence of the Rashba SO coupling
on quantum magnetic oscillations as done below in Sec.~\ref{sec:DOS-oscillations}. \\
(ii) One can derive a simple expression for the correction to the unperturbed spectrum (\ref{LL-(R=0)})
when the values of the parameters $\Delta, B$ are specially adjusted, e.g. in the vicinity of the anticrossing point.
%{\bf Do we need this explanation?}

\subsubsection{The $\Delta =0$ case}
\label{sec:LL-Rashba-Delta=0}

As we already mentioned, the induced by the on-site potential difference between
sublattices gap $\Delta$ can be tuned by adjusting the electric field $E_z$. Thus it is possible to realize
the $\Delta =0$ case. Then the last term of Eqs.~(\ref{f(E,R)=0}) and (\ref{low f(E,R)=0})
becomes zero making these equations biquadratic. Then
the energies of the Landau levels are still given by Eq.~(\ref{LL-(R=0)}),
but the Landau scale (\ref{Landau-scale}) has to be replaced by the renormalized scale $\hbar \tilde \omega$
defined in Eq.~(\ref{renormalized-scale}).

It is convenient to express the renormalized Landau scale $\hbar \tilde \omega = \sqrt{2} \hbar {\tilde v}_F/l_B$
in terms of the renormalized Fermi velocity ${\tilde v}_F^2 = v_F^2 + v_R^2$, where we introduced
the velocity $v_R = \Delta_R a/\hbar$ associated with the Rashba coupling.
Using the characteristic values of the lattice constant
$a = 3.86 \mathring{A}$ and Fermi velocity $v_F = 5.5 \times 10^5 \, \mbox{m/s}$
for silicene, and assuming that the Rashba term is $\Delta_R =\kappa [\mbox{meV}]$,
one can estimate the ratio $v_R/v_F = \kappa  1.1 \times 10^{-3}$.
This indicates that  the impact of the Rashba term is rather small in the considered case
unless $\Delta_R$ is really large.

\subsubsection{Landau levels in the quasiclassical regime}

In the quasiclassical limit, $n \gg 1$ one can neglect the difference between the factors $\sqrt{n}$ and
$\sqrt{n+1} \approx \sqrt{n}$ in the matrix Hamiltonian (\ref{Hamiltonian-matrix-B}).
Then the corresponding characteristic equation acquires the form
\begin{equation}
\label{characteristic-n>>1}
\begin{split}
&\epsilon ^4-2 \left(\Delta ^2+ \Delta _{\text{SO}}^2 + n \hbar ^2\tilde\omega^2 \right)\epsilon^2 +\\
&\left(\Delta_{\xi\uparrow}^2+n\hbar ^2 \tilde\omega^2\right) \left(\Delta_{\xi\downarrow}^2+n \hbar ^2 \tilde\omega^2 \right)
-4n\Delta^2R^2 =0,
\end{split}
\end{equation}
where  we used the renormalized Landau scale (\ref{renormalized-scale})
and separated the last $\sim R^2$ term. One can notice that Eq.~(\ref{characteristic-n>>1})
follows from the general equation (\ref{f(E,R)=0}) if in addition to $n \gg 1$ one assumes that
$|\epsilon|, |\Delta _{\text{SO}}| \ll n |\Delta|$.
It is easy to solve the biquadratic equation (\ref{characteristic-n>>1}) to obtain the large $n$ spectrum
\begin{equation}
\label{LL-large-n}
\begin{split}
& \epsilon _{n \xi \sigma}^{\pm}=\\
& \pm \sqrt{\Delta ^2+\Delta _{\text{SO}}^2+n\hbar^2 \tilde \omega ^2-2 \xi s_\sigma \Delta \sqrt{\Delta _{\text{SO}}^2+n R^2}}.
\end{split}
\end{equation}
Here the factor $\xi s_\sigma$ guarantees that for $R=0$ the spectrum (\ref{LL-large-n}) agrees with Eq.~(\ref{LL-(R=0)})
for $n>0$.

The spectrum (\ref{LL-large-n}) also follows from the Lifshitz-Onsager quantization condition
for  the cross-sectional area of the orbit in momentum space,
\begin{equation}
\label{Onsager}
S(\epsilon) = (n+\gamma) 2 \pi \hbar \frac{e B}{c}.
\end{equation}
One can check this rewriting the area of the orbit $S(\epsilon) = \pi \hbar^2 k^2$ via the momentum
$\hbar \mathbf{k}$ expressed from the inverse zero field dispersion relationship (\ref{spectrum-B=0}).
Then assuming that the phase $\gamma =0$, as it should be in the case of the massive Dirac fermions
\cite{Sharapov2004PRB} (see also Refs.~\cite{Carmier2008PRB,Fuchs2010EPJB}, where the role of the semiclassical
``Berry-like'' phase is studied), and solving Eq.~(\ref{Onsager}) with respect to the energy
$\epsilon$ one reproduces the large $n$ spectrum (\ref{LL-large-n}).

\subsubsection{Energies of the Landau levels near the level anticrossing points}

The points of level crossing are shown in Fig.~\ref{fig:1} computed for
the $\Delta_R =0$ case. They are determined by the
condition $\epsilon_{n \xi \sigma}=\epsilon_{n^\prime \xi \sigma^\prime}$, where the energy
$\epsilon_{n \xi \sigma}$ is given by Eq.~(\ref{LL-(R=0)}).
Then the level crossing condition acquires the form \cite{Ezawa2012JPSJ}
\begin{equation}
(n-n^\prime)\hbar^2\omega^2=2\xi (s_\sigma-s_{\sigma^\prime})\Delta\Delta_{\text{SO}}.
\end{equation}
Accordingly we obtain that the branches with the opposite spin, $s_\sigma^\prime = - s_\sigma$,
cross when the value of the gap is
\begin{equation}
\label{crossing}
\Delta = \pm (n-n^\prime) \Delta^{\mathrm{cros}}, \qquad
\Delta^{\mathrm{cros}} = \frac{\hbar^2 \omega^2}{4\Delta _{\text{SO}}}.
\end{equation}
Comparing Figs.~\ref{fig:1} and \ref{fig:2} we saw that for $R \neq 0$
the anticrossing occurs only at the points $\Delta = - \xi \Delta^{\mathrm{cros}}$.

As shown above the energy $\epsilon_{0\xi\downarrow}$ of the lowest
Landau level given by Eq.~(\ref{LLL-energy}) does not depend on $B$ both for
$R =0$ and $R \neq 0$. The energies of the other low lying levels are
determined by Eq.~(\ref{low f(E,R)=0}). For $R=0$ its solutions are given by Eqs.~(\ref{cubic-root1})
and (\ref{cubic-root23}). As we already saw, the first root is shown in Fig.~\ref{fig:1}
as the solid (blue) straight lines and the other two roots are given by the dashed (red) parabolas
with the lowest absolute value of the energy. It is easy to check that at the energies
$E_{n \xi  \sigma}^{\pm} \equiv \epsilon_{n \xi  \sigma}^{\pm}(\Delta = - \xi \Delta^{\mathrm{cros}})$
there is indeed level crossing, viz.
\begin{equation}
\label{E0-crossing}
E_{0 \xi  \uparrow}= E^{+}_{1 \xi \downarrow}= %-\tilde \epsilon^{-}_{1\xi \downarrow} =
\Delta_{\text{SO}}+\frac{\hbar^2\omega^2}{4\Delta_{\text{SO}}}.
\end{equation}

To estimate how the presence of Rashba term, $R \neq 0$ changes these energies
when level crossing switches to anticrossing, one seeks a solution of Eq.~(\ref{low f(E,R)=0}) in
the following form,  $\epsilon(R )=\epsilon(R=0)+\delta\epsilon$. The corresponding equation for an energy perturbation
$\delta\epsilon$ is
\begin{equation}
\label{delta-E3}
\delta\epsilon^3+2\Delta_{\xi\uparrow} \delta\epsilon^2-(\hbar^2\omega^2+4\xi\Delta\Delta_{\text{SO}}+R^2)\delta\epsilon+2\xi\Delta R^2=0.
\end{equation}
For $\delta E = \delta \epsilon (\Delta = - \xi \Delta^{\mathrm{cros}}) $ Eq.~(\ref{delta-E3}) acquires the form
\begin{equation}
2\Delta_{\text{SO}}\delta E^3+(\hbar^2\omega^2+4\Delta_{\text{SO}}^2) \delta E^2-
2\xi\Delta_{\text{SO}}R^2 \delta E -\hbar^2\omega^2R^2=0.
\end{equation}
We found that the relevant solution of the last equation can be approximated by the following linear in $R$
expression
\begin{equation}
\label{shift1}
\delta E
\approx  \pm \frac{\hbar \omega R}{\sqrt{\hbar^2\omega^2+4\Delta_{\text{SO}}^2}}
\end{equation}
that describes the energy shift of the  crossing  levels that for $R=0$ had the energy (\ref{E0-crossing}).
Taking into account that the energy gap between the anticrossed levels corresponds to the doubled
level shift $\delta E$, one can check that
for $\hbar \omega \gg \Delta_{\text{SO}}$ Eq.~(\ref{shift1})  reduces to Ezawa's result \cite{Ezawa2012JPSJ}
\begin{equation}
\label{Ezawa-shift}
2 \delta E \approx 2 R = 2 \sqrt{2} \frac{a}{l_B} \Delta_{R}.
\end{equation}

Now we pass to the  higher Landau levels with the energies determined by Eq.~(\ref{f(E,R)=0}). For $R=0$
its solutions are given by Eq.~(\ref{LL-(R=0)}). Accordingly, we find that at the level anticrossing point,
$E_{n \xi  \sigma}^{\pm} = \epsilon_{n \xi  \sigma}^{\pm}(\Delta = - \xi \Delta^{\mathrm{cros}})$
the energy is
\begin{equation}
\label{En-crossing}
E_{n  \xi \sigma}^{\pm}= \pm \sqrt{\Delta_{\text{SO}}^2+\frac12(2n+s_\sigma)\hbar^2\omega^2+\frac{\hbar^4\omega^4}{16\Delta_{\text{SO}}^2}}.
\end{equation}
One can see  that for the adjacent levels $E_{n \xi \uparrow}^{\pm}=E_{(n+1) \xi \downarrow}^{\pm}$, so that
for $n=0$ the positive branch of the spectrum reduces to Eq.~(\ref{E0-crossing}).
As in the previous case, we seek for a solution
of the equation (\ref{f(E,R)=0}) at the anticrossing point, $\Delta = - \xi \Delta^{\mathrm{cros}}$,
in the form  $\epsilon(R )=\epsilon(R=0)+\delta\epsilon$ with the perturbation $\delta\epsilon$ caused by a finite $R$.
Neglecting  $\delta\epsilon^4$ term, we
found its approximate solution:
\begin{equation}
\label{shift2}
\begin{split}
 & \delta E \approx  \pm \frac{\hbar\omega R}{2\sqrt{2}\Delta_{\text{SO}} E_{n \xi \sigma}^{+}}\\
& \times \sqrt{ \Delta_{\text{SO}}\left(\Delta_{\text{SO}}+
E_{n \xi \sigma}^{+} \right) + \frac{1}{2}
(2n+ s_\sigma)\frac{\hbar^2\omega^2}2}.
\end{split}
\end{equation}
This expression represents one of the main results of the present work.
Taking $n=0$ and $\sigma = \uparrow$ in Eq.~(\ref{shift2}) one can verify that it reduces to the
derived above Eq.~(\ref{shift1}).

\begin{figure}[ht]
\includegraphics[width=8cm]{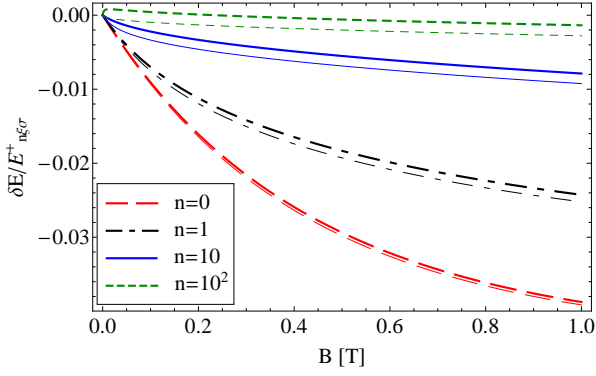}
\caption{ (Color online)
The dependence of the relative energy shift $\delta E/E_{n \xi \sigma}^{+}$
at $\mathbf{K}_{-}$ point as a function of  magnetic field $B$ for
$\Delta_{\text{SO}}=10 \, \text{meV}$ and $\Delta_R=50 \, \text{meV }$.
The long-dashed (red) line is for $n=0$,
$n=1$ -— the dash-dotted (black) line, $n=10$ -- the solid (blue) line, and $n=100$ -- the short-dashed (green) line.
All thick lines are plotted using the energy difference
$\delta E=\epsilon_{n \xi}(R\ne0,\Delta = - \xi\Delta^{\mathrm{cros}})- \epsilon_{n \xi}(R=0,\Delta = - \xi\Delta^{\mathrm{cros}})$
which is computed using the numerical solution of the general Eq.~(\ref{f(E,R)=0})
and the thin lines are calculated using the approximate Eq.~(\ref{shift2}). }
\label{fig:4}
\end{figure}

\begin{figure}[ht]
\includegraphics[width=8cm]{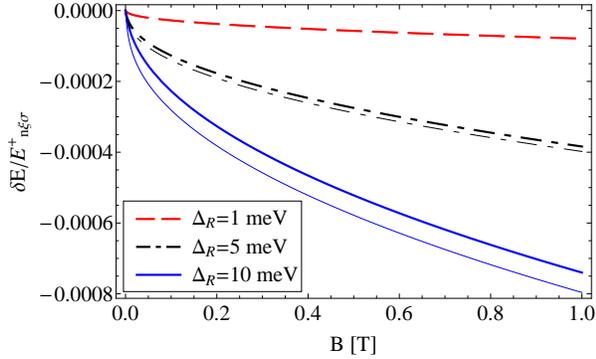}
\caption{ (Color online)
The dependence of the relative energy shift $\delta E/E_{n \xi \sigma}^{+}$
at $K_{-}$ point as a function of  magnetic field $B$ for
$\Delta_{\text{SO}}=10 \, \text{meV}$ and $ n= 50$.
The dashed (red) line is for $\Delta_R=1 \, \mbox{meV}$ ,
$\Delta_R=5 \, \mbox{meV}$  --
the dash-dotted (black) line, and
$\Delta_R=10 \, \mbox{meV}$
-- solid (blue) line.
All thick lines are plotted using the energy difference
$\delta E=\epsilon_{n \xi}(R\ne0,\Delta = - \xi\Delta^{\mathrm{cros}})- \epsilon_{n \xi}(R=0,\Delta = - \xi\Delta^{\mathrm{cros}})$
which is computed using the numerical solution of the general Eq.~(\ref{f(E,R)=0})
and the thin lines are calculated using the approximate Eq.~(\ref{shift2}).
}
\label{fig:5}
\end{figure}

We plot in Figs.~\ref{fig:4} and \ref{fig:5} the exact result based on the numerical solution
of Eq.~(\ref{f(E,R)=0}) and the approximate expression (\ref{shift2}) to investigate the range of its validity.
Figure~\ref{fig:4} shows the dependence of the relative energy shift $\delta E/E_{n \xi \sigma}^{+} =
[\epsilon_{n \xi}(R\ne0,\Delta = - \xi\Delta^{\mathrm{cros}})- \epsilon_{n \xi}(R=0,\Delta = - \xi\Delta^{\mathrm{cros}}) ]/E_{n \xi \sigma}^{+}$ at $\mathbf{K}_{-}$ point as a function of  magnetic field $B$
for a fixed value of $\Delta_R$ and four different values of $n =0, 1, 10, 100$.
The values $\Delta_{\text{SO}}=10 \, \text{meV}$ and $\Delta_R=50 \, \text{meV }$ are taken.
The thick lines are plotted using the energy difference
$\delta E=\epsilon_{n \xi}(R\ne0,\Delta = - \xi\Delta^{\mathrm{cros}})- \epsilon_{n \xi}(R=0,\Delta = - \xi\Delta^{\mathrm{cros}})$
which is computed using the numerical solution of the general Eq.~(\ref{f(E,R)=0})
and the thin lines are calculated using the approximate Eq.~(\ref{shift2}).
We observe that the expression for $\delta E$ provides rather good approximation for the energy
shift at the anticrossing point for all values of $n$ and even for a large value of $\Delta_{R}$.
%As $n$ increases the

In Fig.~\ref{fig:5} we plotted the dependence of the relative energy shift $\delta E/E_{n \xi \sigma}^{+}$ at $\mathbf{K}_{-}$ point
as a function of  magnetic field $B$ for three values of $\Delta_R=1 \, \mbox{meV}$,
$\Delta_R=5 \, \mbox{meV}$, and $\Delta_R=10\, \mbox{meV}$  for fixed $n=50$ and
$\Delta_{\text{SO}}=10 \, \text{meV}$.
We also observe that for a small value of $\Delta_{R}$ the approximated expression practically coincides with the
exact one. As $\Delta_{R}$ increases, the approximate result deviates from the exact one. This is
not surprising, because the expression (\ref{shift2}) was obtained using a linear in $R$ approximation.

\section{The density of states}
\label{sec:DOS-oscillations}

In the absence of scattering from impurities the  density of states (DOS) is expressed via
the energies of the Landau levels $\epsilon_{n\xi \sigma}^\lambda$ as follows:
\begin{equation}
\label{DOS-def}
D_0(\varepsilon)=\frac{eB}{2\pi \hbar c}
\sum_{\substack{\sigma =\downarrow , \uparrow \\ \xi = \pm } } \sum_{\lambda = \pm}
 \sum_{n=0}^{\infty} \delta(\varepsilon-\epsilon_{n\xi \sigma}^\lambda).
\end{equation}
The broadening of Landau levels due to the scattering from impurities can be taken into account \cite{1984ShonbergBook}
by convolution of the DOS  $D_0(\omega)$ with zero level broadening
with the distribution function $P_\Gamma(\omega)$, viz.
\begin{equation}
\label{level-broad}
D(\varepsilon)=\int\limits_{-\infty}^\infty d\omega P_{\Gamma}(\omega-\varepsilon)D_0(\omega), \qquad \int\limits_{-\infty}^\infty d\omega P_{\Gamma}(\omega)=1.
\end{equation}
The simplest model for the level broadening (\ref{level-broad}) is the Lorentz distribution,
$P_\Gamma(\omega) = \Gamma/\left[\pi(\omega^2+\Gamma^2)\right]$
with the impurity scattering rate $\Gamma$.

In its turn the knowledge of zero temperature DOS is completely sufficient to write down
the finite temperature thermodynamic potential and other thermodynamic quantities.
Moreover, the DOS can be experimentally found by measuring the
quantum capacitance $C$ \cite{Ponomarenko2010PRL,Yu2013PNAS}, which is
proportional to the thermally smeared DOS and is given by
\begin{equation}
\label{capacitance}
C(\mu )=e^{2}\int\limits_{-\infty }^{\infty }d\varepsilon D(\varepsilon )\left(
-n_{F}^{\prime }(\varepsilon )\right) ,
\end{equation}%
where $n_{F}(\varepsilon )=1/[\exp (\varepsilon -\mu )/T+1]$ is the Fermi
distribution.

\subsection{The DOS in the absence of Rashba term}

As was discussed in the Introduction, the electron subsystem
in the considered Dirac materials for $\Delta_R =0$ turns out to be equivalent to
two independent layers of the gapped monolayer graphene
with the gaps $\Delta  \pm \Delta_{\text{SO}}$. Indeed, the $R=0$ spectrum
(\ref{LL-(R=0)}) for a fixed value of the spin $\sigma$ reduces to the well-known spectrum of the
gapped graphene \cite{Haldane1988PRL} (see also Appendix~D of Ref.~\cite{Gusynin2006PRB}). Thus using the results of
Ref.~\cite{Sharapov2004PRB} one can straightforwardly write the final expressions for the DOS.

In the absence of scattering from impurities using the Poisson summation formula
one can  derive from Eq.~(\ref{DOS-def}) the following expression:
\begin{equation}
\label{dos1}
\begin{split}
& D_0(\varepsilon)=\frac1{2\pi v_F^2\hbar^2}\sum_{\sigma=\downarrow, \uparrow} \mbox{sgn} (\varepsilon)\frac d{d \varepsilon}  \Bigg(
\theta(\varepsilon^2-\Delta_{+\sigma}^2) \times \\
& \left. \left[ \varepsilon^2-\Delta_{+ \sigma}^2 +
\hbar^2 \omega^2 \sum_{k=1}^{\infty} \frac 1{\pi k}
\sin\left( \frac{2 \pi k (\varepsilon^2-\Delta_{+ \sigma}^2)}{\hbar^2 \omega^2}\right) \right] \right).
\end{split}
\end{equation}
The DOS (\ref{dos1}) contains oscillations with the
two frequencies
$2 \pi k (\varepsilon^2 -\Delta_{+ \downarrow}^2)/(\hbar^2 \omega^2)$  and
$2 \pi k (\varepsilon^2 -\Delta_{+ \uparrow}^2)/(\hbar^2 \omega^2)$.
The oscillatory part of the DOS can be written  in the form  of the beats
\begin{equation}
\label{beats}
\begin{split}
& D_0^{\mathrm{osc}}(\varepsilon)=
\frac{ 2 }{\pi^2 l_B^2}\mbox{sgn} (\varepsilon) \\
& \times
\left[\frac d{d \varepsilon}
\sum_{k=1}^{\infty} \frac1{k}
%\right. \\ & \left.
\sin\left(\frac{2 \pi k F_o}{B}\right)
\cos\left( \frac{2\pi k F_b}{B} \right) \right],
\end{split}
\end{equation}
where
\begin{equation}
\label{B-F}
F_o = \frac{\varepsilon^2 -\Delta_{\text{SO}}^2-\Delta^2}{2 v_F^2 \hbar e/c }
\end{equation}
is the frequency (for $k=1$) of oscillations in $1/B$ and
\begin{equation}
\label{F-b}
F_b=\frac{ \Delta_{\text{SO}}\Delta}{v_F^2\hbar e/c}
\end{equation}
is the frequency (for $k=1$) of beats. In deriving Eq.~(\ref{beats})
we assumed that $|\varepsilon |> |\Delta \pm  \Delta_{\text{SO}}|$.
For $\varepsilon^2 \gg |\Delta_{\text{SO}} \Delta|$ the frequency of beats
$F_b \ll F_o$. Notice that in the absence of the Rashba interaction  the frequency $F_b$
depends solely on the SO gap $\Delta_{\text{SO}}$ and is tunable by the applied electric field gap $\Delta$.

In the case of the distribution function $P_{\Gamma}(\omega)$  given by the Lorentzian distribution,
the sum over Landau levels can be
expressed in the closed form \cite{Sharapov2004PRB} in terms of the digamma function $\psi$.
Accordingly, the DOS of silicene represents the sum of the two terms
\begin{equation}
\label{dos-dirty}
\begin{split}
& D(\varepsilon)=\frac{1}{\pi^2 v_F^2\hbar^2}\left\{2\Gamma\ln\left(\frac{\Lambda^2}{2\hbar v_F^2 eB/c}\right)
-  \mbox{Im} \left[(\varepsilon+i\Gamma)\times \right. \right. \\
&  \left. \left. \sum_{\sigma=\downarrow, \uparrow}  \left(\psi\left(\frac{\Delta_{+ \sigma}^2-(\varepsilon+i\Gamma)^2)}{2\hbar v_F^2 eB/c} \right)+\frac{\hbar v_F^2 eB/c}{\Delta_{+ \sigma}^2-(\varepsilon+i\Gamma)^2) }\right) \right]\right\},
\end{split}
\end{equation}
where $\Lambda$ is the energy cutoff associated with the bandwidth.
Its presence in the nonoscillatory part of the DOS is related to the Lorentzian shape of the level broadening.

Equation~(\ref{dos-dirty}) turns out to be convenient for numerical modeling of the DOS  when
the width of all Landau levels is the same.
In the case when each level has a different width Eq.~(\ref{DOS-def}) acquires the form
\begin{equation}
\label{DOS-widened}
D_0(\varepsilon)=\frac{1}{2\pi^2 l_B^2}
\sum_{\substack{\sigma =\downarrow , \uparrow \\ \xi = \pm } } \sum_{\lambda = \pm}
 \sum_{n=0}^{\infty}  \frac{\Gamma_{n\xi \sigma}^\lambda}{
 (\varepsilon-\epsilon_{n\xi \sigma}^\lambda)^2 + (\Gamma_{n\xi \sigma}^\lambda)^2}.
\end{equation}
Since the level width $\Gamma_{n\xi \sigma}^\lambda$ is in general unknown, it is impossible to use the
Poisson formula or to sum over Landau levels as done above.
It is possible instead to consider analytically  the thermal smearing of the DOS which is present
in the capacitance (\ref{capacitance}). The final results obtained in Ref.~\onlinecite{Gusynin2014FNT}
can be rewritten as follows
\begin{equation}
\label{capacitance-final}
C(\mu) =
\frac{e^2}{2\pi^2 l_B^2}
\sum_{\substack{\sigma =\downarrow , \uparrow \\ \xi = \pm } } \sum_{\lambda = \pm}
 \sum_{n=0}^{\infty} I(\mu - \epsilon_{n\xi \sigma}^\lambda, \Gamma_{n\xi \sigma}^\lambda),
\end{equation}
where
\begin{equation}
I (\epsilon,\Gamma) = \frac{1}{2 \pi T} \mbox{Re} \psi^\prime \left(
\frac{1}{2} + \frac{\Gamma - i \epsilon}{2 \pi T}\right)
\end{equation}
is expressed in terms of the derivative of the digamma function $\psi$.
The capacitance (\ref{capacitance-final}) already includes both thermal and impurity averages and only the sum over
Landau levels is left for the numerical calculation. Equation~(\ref{capacitance-final}) is in fact valid
not only for $R =0$. In the $R \neq 0$ case instead of the energies $\epsilon_{n\xi \sigma}^\lambda$
given by Eq.~(\ref{LL-(R=0)}) one should use the energies of the corresponding Landua levels found in
Sec.~\ref{sec:LL-Rashba}.

\subsection{The DOS in  the presence of Rashba term}

The expression (\ref{beats}) presented above  can be generalized for the case of $\Delta_{R} \neq 0$.
The analytical expression (\ref{LL-large-n}) valid in the large $n$ limit allows one to evaluate the sum over Landau levels.
First Eq.~(\ref{DOS-def}) can be rewritten as follows:
\begin{equation}
D_0(\varepsilon)=\frac{eB}{\pi \hbar c} \mbox{sgn} \, \varepsilon \frac{d}{d \varepsilon}
\sum_{\sigma = \uparrow, \downarrow} \sum_{n=0}^{\infty} \theta(\epsilon^2 - (\epsilon_{n + \sigma}^{+})^2),
\end{equation}
where the summation over $\lambda, \xi = \pm$ is done.
Then using the Poisson summation formula
\begin{equation}
\begin{split}
& \frac{1}{2}F(0)+\sum\limits_{n=1}^\infty F(n)\\
&=\int\limits_0^\infty
F(x)dx+2{\rm Re}\sum\limits_{k=1}^\infty \int\limits_0^\infty
F(x)e^{2\pi ikx}dx,
\end{split}
\end{equation}
we find that the oscillatory part of the DOS can still be written in the form of Eq.~(\ref{beats}).
The frequency of oscillations
 \begin{equation}
\label{B-F-Rashba}
F_o \approx \frac{\varepsilon^2 -\Delta_{\text{SO}}^2-\Delta^2 + 2 (v_R^2/ v_F^2) \Delta^2 }{2 v_F^2 \hbar e/c }
\end{equation}
is now shifted with respect to its $\Delta_R=0$ value given by Eq.~(\ref{B-F}).
Since the ratio $v_R/v_F$ is small (see Sec.~\ref{sec:LL-Rashba-Delta=0})
and the new term can be absorbed by renormalizing the value of $\Delta$
one can conclude that  this  shift of the oscillation frequency cannot be used to determine the Rashba term.
The situation with the frequency of beats $F_b$ seems to be more promising. Indeed,
Eq.~(\ref{F-b}) acquires the form
\begin{equation}
\label{F-b-Rashba}
F_b=\frac{ \Delta_{\text{SO}}\Delta}{v_F^2\hbar e/c} \left(1+  \frac{1}{2}\frac{v_R^2}{v_F^2} \frac{\varepsilon^2 - \Delta^2 - \Delta_{\text{SO}}^2}{\Delta_{\text{SO}}^2} \right).
\end{equation}
We observe that in the last term in the brackets of Eq.~(\ref{F-b-Rashba}) the smallness of the
ratio $v_R^2/v_F^2 \sim \kappa^2 \times 10^{-6}$
can be compensated by the large value of the ratio $\varepsilon^2/ \Delta_{\text{SO}}^2$.
Even more important is that due this term the frequency $F_b$ is now dependent on  the position of the Fermi level,
$\varepsilon = \mu$, and, accordingly, on the carrier concentration.

\section{Conclusion}
\label{sec:concl}

We studied how the pattern of Landau levels in
the low-buckled Dirac materials is modified by
the intrinsic Rashba SO coupling between NNN.
In particular, we  found the approximate analytical expressions (\ref{shift1}) and (\ref{shift2})
for the energy shift caused by the Rashba term in the vicinity of the level anticrossing points.
The impact of the Rashba interaction is maximal in this regime.

We also derived the analytical expression (\ref{LL-large-n}) for energies of the
Landau levels in the large $n$ limit. Its relatively simple form allowed us to derive the analytical expression
describing quantum magnetic oscillations of the DOS. A specific feature of the oscillations  is
the presence of the beats caused by crossing of the Fermi level by the Landau levels
from the two different branches of the quasiparticle excitations. These beats resemble the oscillatory effects
observed in the usual 2D electron gas with parabolic dispersion and
Rashba interaction \cite{Bychkov1984JPCS}. When the Rashba interaction is absent, the frequency of beats $F_b$ is
given by Eq.~(\ref{B-F}). It is proportional to the product $\Delta_{\text{SO}}\Delta$, where
the sublattice asymmetry gap $\Delta$ can be controlled by the applied electric field $E_z$.
In the presence of the intrinsic Rashba interaction  the frequency $F_b$ shifts (\ref{B-F-Rashba})
and becomes dependent both on the gap $\Delta$ and carrier concentration. This peculiarity can be helpful
for the experimental determination of the value of the Rashba coupling constant.

Our results are applicable in the analysis of a number of experiments which probe transport and thermodynamic
properties of the  low-buckled Dirac materials, including cyclotron resonance, tunneling spectroscopy,
capacitance measurements, charge compressibility, and magnetization. Concluding we also note that these results may be
applicable for a wider range of materials, e.g., for a bilayer TI \cite{Zhang20013PRL}.

\begin{acknowledgments}
S.G.Sh gratefully acknowledges E.V.~Gorbar, V.P.~Gusynin and V.M.~Loktev for helpful discussions.
The authors acknowledge the support of the
European IRSES Grant SIMTECH No. 246937.
%S.G.Sh. was also supported by SFFR of Ukraine, grant No.~F53.2/028.

\end{acknowledgments}

\end{document}